\documentclass{llncs}
\input{psfig.sty}
\usepackage{llncsdoc}
\usepackage{ifpdf}
\usepackage{amsmath}
\usepackage{graphicx}
\DeclareGraphicsExtensions{.jpg}
\usepackage[ansinew]{inputenc}

\begin{document}

\frontmatter          

\pagestyle{headings}  

\mainmatter              
\title{A Flexible Structured-based Representation for XML Document Mining}
\titlerunning{XML Document Mining}  
%
\author{Anne-Marie Vercoustre \and Mounir Fegas \and Saba Gul 
 \and Yves Lechevallier}
\authorrunning{Anne-Marie Vercoustre et al.}   
\institute{INRIA, Rocquencourt, France\\
\email{Firstname.Lastname@inria.fr},\\ WWW home page:
\texttt{http://www.inria.fr/index.en}
}
\maketitle              

\begin{abstract}
This paper reports on the INRIA group's approach to XML mining while participating in the INEX XML Mining track 2005. 
We use a flexible representation of XML documents that allows taking into account the structure only or both the structure and content. 
Our approach consists of representing XML documents by a set of their sub-paths, defined according to some criteria (length, root beginning, leaf ending).
By considering those sub-paths as words, we can use standard methods for vocabulary reduction, 
and simple clustering methods such as $k$-means. 
We use an implementation of the clustering algorithm known as
\textit{dynamic clouds} that can work with distinct groups of independent modalities put in 
separate variables. This is useful in our model since embedded sub-paths are not 
independent: we split potentially dependant paths into separate variables, resulting 
in each of them containing independant paths.
Experiments with the INEX collections show good results for the structure-only collections, but our approach could not scale well for large structure-and-content collections.

\end{abstract}
\section{Introduction}

XML documents are becoming ubiquitous because of their rich and
flexible format that can be used for a variety of applications.
Standard methods have been used to classify XML documents, reducing
them to their textual parts~\cite{doucet02naive}. These approaches do not take advantage of
the structure of XML documents that also carries important information.

Recently much attention has been drawn towards using the structure of XML documents to improve information retrieval, 
classification and clustering, and more generally information mining~\cite{costa04tree,dalamagas04structure,denoyer04apprentissage,guillaume00clustering,termier02treefinder}. In the last four years, 
the INEX (Initiative for the Evaluation of XML retrieval) has focused on system performance 
in retrieving elements of documents rather than full documents and evaluated the benefits for end users.
Other researches have focussed on clustering large collections of documents using representations of documents that involve both the structure and the content of documents, or the structure only~\cite{Denoyer05,vfld06a,yoon01bitcube}. One motivation for structured-based clustering is to organise XML
collections into smaller collections with a specific schema that supports optimisation of the query processing.

Approaches for combining structure and text range from adding a flat representation of the structure 
to the classical vector space model~\cite{doucet02naive} or combining different classifiers for different tags or media~\cite{denoyer03}, 
to defining a more complex structured vector models~\cite{yi00classifier}, possibly involving attributes and links~\cite{jianwu02}.

When using the structure only, the objective is generally to organize large and heterogeneous collections 
of documents into smaller collections (clusters) that can be stored and searched more effectively.  
Part of the objective is to identify substructures that characterize the documents in a cluster and 
to build a representative of the cluster~\cite{flesca02detecting}, possibly a schema or a DTD.

Since XML documents are represented as trees, the problem of clustering XML documents can be seen as the same as clustering trees. 
One can identify two main approaches: 1) identify frequent common sub-patterns between trees and group 
together documents that share the same patterns~\cite{termier02treefinder,zaki03xrules,flesca02detecting}; 
2) define a similarity measure between trees that can be used with a standard clustering algorithm. 
A possible distance is calculated by associating a cost function to the edit distance between two trees~\cite{francesca03distance,nierman02similarity,dalamagas04structure}. 
However, it is well known that edit distance algorithms have complexity issues. 
Therefore some models replace the original trees by structural summaries~\cite{dalamagas04summaries} or s-graphs~\cite{lian04} 
that retain only the intrinsic structure of the tree: for example, reducing a list of elements to a single element, 
flattening recursive structures, etc. 

A common drawback of the two approaches above is that they reduce documents to their intrinsic patterns (sub-patterns, 
or summaries) and do not take into account an important characteristic of XML documents: the notion of list of elements and more precisely the number of elements in those lists.  
While it may be fine for clustering heterogeneous collection, suppressing lists of elements may result 
in losing document properties that could be interesting for other types of XML mining.

Our idea is therefore to use a document representation that takes into account the frequency of structure within the documents, 
while not be as costly as the edit distance.

In this paper we propose a generic model for clustering documents that involves either their structure 
or both their structure and content. 
We represent documents by flattening their trees into their sets of sub-paths of length between $n$ and $m$, 
two \textit{a priori} given values. 
We retain the frequency of paths and we consider sub-paths as words.
Therefore we can apply standard clustering methods usually used for text. 
When considering document content as well as structure, sub-paths are extended with the individual words of the text contained 
in the terminal node of each path.
For specific values of $m$ and $n$, our model is equivalent to models that have been proposed before, so we offer a more general framework.

 We evaluate our model using the collections proposed in the INEX XML mining track, 
while being aware that our approach may not be appropriate for some of the proposed collections, 
in particular those where the order of elements is significant for clustering. 

In Sect. 2, we present our document model for clustering and compare it, in Sect. 3, 
to previous models for specific values of $m$ and $n$. Sect. 4 describes our clustering method and some additional feature selection. 
Sect. 5 details the evaluation metrics we use, while Sect. 6 and 7 present our experiments and the results. In Sect. 8 we propose our conclusions.

\section{Our Model for Document Representation}

XML documents are usually represented as trees where each node corresponds to an XML tag. 
The hierarchy of the nodes reflects the  embedding of the tags, and leaf nodes have associated text. 
Attributes can be represented the same way as sub-elements, i.e. as additional descendants 
of the node they are attached to. 

\begin{figure}
\ifpdf
	\centering
	\includegraphics[height=8cm]{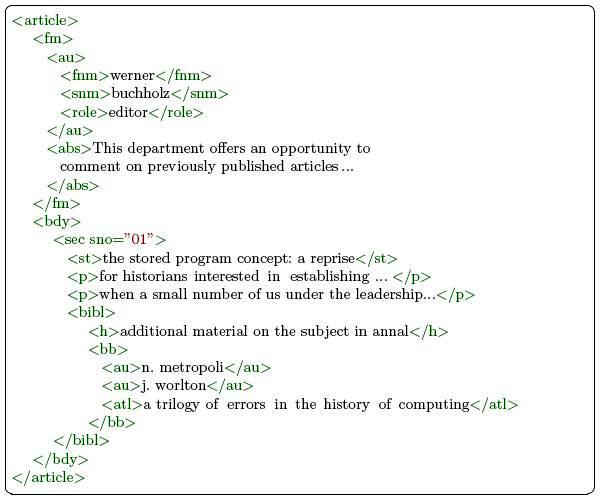}
\else
	\centerline{\psfig{figure=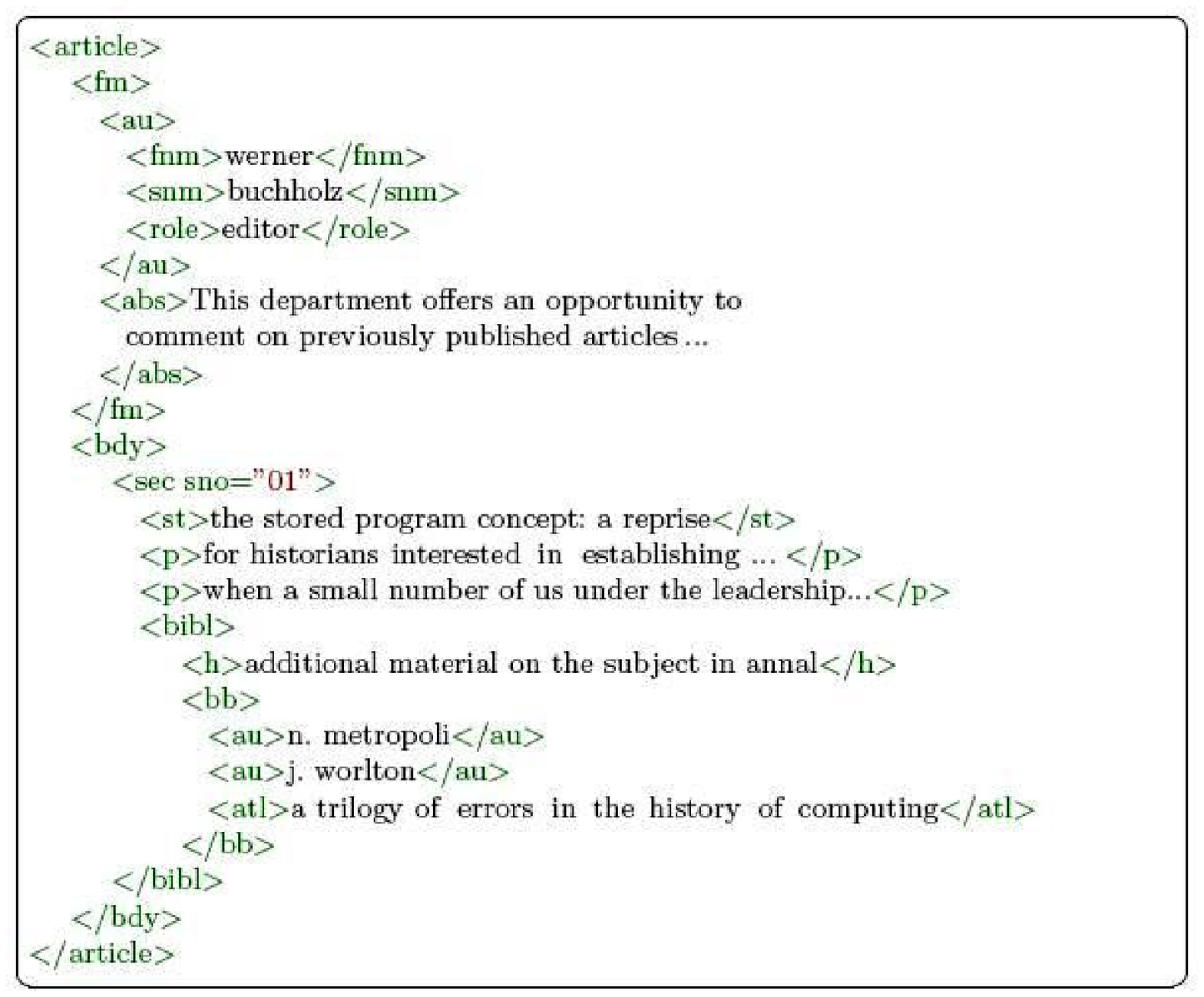,height=8cm}}
\fi
\caption{An example of XML document}
\label{fig:document}
\end{figure}

Fig.~\ref{fig:document} gives an example of an XML document, extracted from the IEEE collection,
that we will use throughout the paper.

Our model is based on a tree linearization that represents a document as its set of paths. 
The precise definition of paths to consider is  defined below and correspond, in fact, to a family of possible representations 
that take into account the structure, the text, or both.
By regarding paths as simple words we can use the vector model to represent documents from their structure.

The motivation for a flexible choice of paths or sub-paths in the document is 
that some analysis or clustering tasks 
may be interested in the top part of the tree, the lower parts of the tree, or possibly parts in the middle. 
An example would be clustering very heterogeneous collections based on the structure, 
where the partition can be done by looking at the top level elements only. On the contrary, 
if one wants to cluster documents based mostly on the text, it could be appropriate to add some limited context just above the text.

Before presenting our structured document representations, we introduce some definitions:

\begin{definition}
The \it{path of a node} $n$ is the sequence of nodes from the root to this node, when traversing the tree from child to child. 
We note it $p(n)$. 
It is also called a \it{root-beginning path}\footnote{We use the terminology used in Liu and \textit{al.}~\cite{liu04pca} for complete paths, 
root-beginning paths, leaf-ending paths}, or \it{root path} for short.
\end{definition}

\begin{definition}
The {\it length of a path} is the number of nodes in the path.
\end{definition}

\begin{definition}
A \emph{sub-path s of length l} on a path p is a sequence of l consecutive nodes along the path p. 
(i.e. a sub-path does not necessarily start at the root). We note $|s|$ the length of  the sub-path s. 
\end{definition}

Table~\ref{tab:path1} shows examples of paths and sub-paths of length 3.

To take into account the text of the documents, we introduce ``text paths" defined as follow:

\begin{definition}
A {\it text sub-path} is a sub-path that ends with a word contained in 
the text associated with the last node in the sub-path. When the last node is not a leaf, the words are those associated with its descendants.
\end{definition}

\begin{table}
\begin{minipage}{0.47\linewidth}

\centering

\tabcolsep = 4\tabcolsep

\begin{tabular}{lc}

\hline\hline

Paths&Tf\\ 

\hline

article.bdy.sec & 1\\

article.fm.au & 1\\

bdy.sec.p & 2\\

bdy.bb.au & 2\\

bdy.sec.sno@ & 1\\ \hline
\\
\end{tabular}

\caption{Paths and sub-paths of length 3; the character {\upshape @} marks an attribute.}

\label{tab:path1}

\end{minipage} \hfill
\begin{minipage}{0.47\linewidth}

\centering

\tabcolsep = 4\tabcolsep

\begin{tabular}{lc}

\hline\hline

Paths&Tf\\ \hline

article.fm.abs.``offer" & 1 \\

bdy.sec.p.``historian" & 1 \\

bdy.sec.sno@.``01" & 1 \\ 

article.fm.au.``werner" & 1 \\

bdy.sec.``historian" & 1 \\

\hline
\\
\end{tabular}

\caption{Textual paths of length 4 and 3.}

\label{tab:path2}

\end{minipage}

\end{table}

Table~\ref{tab:path2} shows some text paths or sub-paths of length 4 and 3 
corresponding to the example in Fig.~\ref{fig:document}. 
The last two paths are non terminal paths extended with words that 
are not directly associated  with their final node but with one of their descendants.

We can now define a family of representations for a XML document tree $d$ as:
\begin{equation} \label{eq:F}
R(d)=\sum_{i}{w_i p_i}
\end{equation}
 for all  sub-paths $p_i$  in $d$, where $m  \le | p_i | \le  n , 1 \le  m \le  n$ , $w_i$ is the frequency of sub-paths $p_i$

The actual representations are defined by a few parameters:
\begin{itemize}
\item	$m$ and $n$ are two  \textit{a priori} fixed integers. The value ``$n$" can be replaced with the symbol ``*", 
meaning that, for each sub-path, the maximum value would be the length of its supporting path.
\item	when the parameter {\bf root} is set on,  only the sub-paths starting from the root (root-beginning paths) are generated.
\item	when the parameter {\bf leaf} is set on, only the sub-paths ending at leaf nodes (leaf-ending path) are generated.
\item	with the parameter {\bf text} set on, only ``text paths" are generated.
\item	with the parameter {\bf text-and-node} set on, both  text sub-paths and node sub-paths are generated. 
\item	with the parameter {\bf attribute} set on, attributes, as well as nodes, are considered for path generation. 
\end{itemize}
By setting different parameter values, we can use a variety of document representations for different clustering tasks.
Before presenting our clustering approach, we are going to interpret the models for some specific values of the parameters, and compare them  with other existing models.

\section{Comparison with other models for structured documents}

Our document model integrates various representations that have been proposed in other works:

\begin{itemize}
\item	Case $min=1$, $max=1$, $text=true$;

This case corresponds to representing a document by its text only (standard vector model)
\item	Case $min=1$, $max=1$, $text=false$, [$attribute=(false|true)$];

Corresponds to representing a document by the list of its tags.
This is the model used, with or without attributes, in Doucet and \textit{al.}~\cite{doucet02naive}, for the case ``Tag feature only",
  based on the vector model.
It is also used in Flesca and \textit{al.}~\cite{flesca02detecting} where both elements and attribute names are considered.
Moreover if node-and-text is set to true, we get the ``Tag and text features" used in Doucet and \textit{al.}~\cite{doucet02naive}.
\item	Case $min=1$, $max=*$, $root=true$, $leaf=true$, [$text=(false|true)$];

In this case XML documents are represented by the set of their paths from the root to the leaves. 
In Yoon and \textit{al.}~\cite{yoon01bitcube} they use a bitmap matrix where lines represent the documents and columns represent the different terminal paths in the collection. 
The frequency is not used. They also extend their bitmap model by adding quadruplets (document, path, term, $b$) 
where $b$ is true if the path contains the term, which corresponds in our case with text set to true.
When $text=true$, it is also the representation used in Yi and Sundaresan~\cite{yi00classifier} for the ``flat with structured tag" experiments, 
where each term (document word) is replaced by its \emph{text path} in a flat vector. 
\item	Case $min=1$, $max=*$, $root=true$, $leaf=false$, $text=true$; 

A document is represented by the set of all the root text paths (of any length) in the document, 
where a term will belong to its parent node and all the embedding nodes.  
This model is equivalent to the Structure Vector Model proposed in Yi and Sundaresan~\cite{yi00classifier}, 
where a document is represented by all its paths of length 
between 1 and the height $h$ of the document tree. The frequency of terms associated with a path is relative to the subtree associated with that path.
\item   Case $min=1$, $max=L$, $root=false$, $leaf=false$, $text=false$;

One of the representation proposed in Liu and \textit{al.}~\cite{liu04pca} is based on paths of length smaller than $L$, although they can also fix the level in the tree where the paths must start.
In our case paths will start at the root or at any level in the tree.
\item	Case $min=n$, $max=n$, $root=false$, $leaf=true$, $text=true$;

This case corresponds to representing documents by leaf-ending sub-paths of length $n$, 
and therefore providing a limited context to the terms in the documents. 
One of the representations developed in Liu and \textit{al.}~\cite{liu04pca} includes the definitions of leaf-ending paths as well root-beginning paths, of length less than $L$.
They seem to use text as well, but this is not clearly described.
\end{itemize}
Our representation of XML documents using sub-paths is therefore flexible enough to subsume many 
of the representations used in the above works.

Other representations for XML trees for clustering have recently been proposed.
Nayak and Xu~\cite{nayak05} represents an XML document by its level structure: each level is represented by the list of labels (tags) that occur at
this level; multiple instances are ignored and the order of the labels is not preserved. 
The clustering algorithm is based on a similarity measure between levels.
Candillier and \textit{al.}~\cite{candillier05} transforms the XML trees into sets of attribute-values in order to apply various existing methods on such data.
Considered attributes include the set of parent-child relations, the set of next-sibling relations, the set of distinct root paths, etc.
Thoses attributes results in a number of features whose values, for a given document, 
are their number of occurences of this feature in the document. For clustering or classification, 
they use an adaptation of SSC~\cite{SSC05}, a {\it subspace clustering algorithm} 
that has the advantage of providing an interpretable representation of the resulting clusters, 
as a decision tree on the discriminent features.

\section{Clustering Approach}

Since we represent XML documents as a set of paths seen as words, we can use traditional clustering methods for flat texts. 
However we have to deal with two issues: first, reducing the number of paths in case they are too many; 
secondly, the possible dependency of paths. Before presenting the clustering approach we address these two issues.

\subsection{Further feature selection}

Algorithms for clustering such as the $k$-means are linearly dependent on the size of the data, 
that is, the number of words that represent the documents. 
In our case the document will be represented not only by the different words in the text but possibly by their contextual paths 
(i.e. generating as many different occurrences of a word as the different contexts in which it occurs). 
Moreover they may be extra ``words" corresponding to any sub-path in the document trees (node paths)\footnote{
Obviously there will be much more leaf-ending paths than root-paths since trees are expending from the root to the leaves}.
It is therefore necessary to limit the number of paths that represent the documents to reduce the clustering time. 

We apply two levels of feature selection in the path generation: structure level and text level. 
Then we reduce the number of paths by applying standard selection on words using their relative frequency (TF/IDF).

\paragraph{{\bf Structure level}}
Usually we reduce the number of generated paths by regrouping some tags in more semantic categories, 
using our knowledge of the DTD or the collections. 
For example we replace tags for different sections ($ss1$, $ss2$, $sec$), by a single tag ``$sec$", and ignore presentation tags.  
Since the INEX collections were specially preprocessed for the XML mining track,  we did not have to take care of these semantic groupings. 

\paragraph{{\bf Text level}}
For the textual content of the document, we use standard reduction methods:
\begin{itemize}
\item  stop list word for suppressing insignificant word
\item suppression of  terms shorter than 4 characters
\item pseudo stemming using the Porter stemmer~\cite{porter97}.
\end{itemize}
	 
\paragraph{{\bf Frequency of paths}}
 
As said before, the documents are represented by a set of paths that depends on the chosen parameters. 
First, the frequency of a path is calculated and normalized using the TF/IDF formula
(number of path occurrences in the document over their number in the whole collection).
Paths that are too frequent or too rare will be suppressed. 
In particular paths that occur only once in the collection will be suppressed since it will not affect the clustering process. Similarly paths that occur in every document will not contribute to partitioning documents.

For the remaining paths, we calculate their normalised weight in the document by dividing the number of occurrences 
of the path by the number of the paths in the document (standard vector normalization to take into account the length of the documents).

\subsection{Word Dependency}

Clustering algorithms based on the vector model rely on the independence of the various dimensions (words) 
for calculating the distance between the vectors. 
Although this is not always verified in practice with words in texts, it usually works fine. 
In our case, where words are sub-paths in the document tree, there is an obvious dependency between embedded sub-paths.
For example, the two paths $bdy.sec$ and $bdy.sec.st$ are not independent since the second can exist only 
if the first one exists, 
the first one being embedded in the second one. However, two overlapping paths, such as $bdy.sec$ and $sec.st$ would not be regarded as dependent.   
We only consider structural dependency  here, not dependency that would derive from the DTD itself, 
for example if two siblings are mandatory according to the DTD definition. 
This later dependency would not affect the clustering results since the two paths would then be present in all the documents and 
therefore eliminated as very frequent.

To deal with the first case of dependency, we partition the paths by their length and treat each set of paths 
as different variables in the clustering algorithm as explained below. 

\subsection{Clustering Method}

Our clustering algorithm is based on the partitioning method proposed by Celeux and \textit{al.}~\cite{celeux89}, where the distance between clusters is based on the frequency of the words of the selected vocabulary. 
This approach is equivalent to the $k$-means algorithm. As for the $k$-means we  
represent the clusters by prototypes which summarize the information (paths) of the documents belonging to each of them.

More precisely, if the vocabulary counts $p$ words, each document $s$ is represented by the vector 
$x_s=(x_s^1,...,x_s^j,...,x_s^p)$ where $x_s^j$ 
is the number of occurrences of word $x_j$ in the document $s$, then the prototype $g$ for a class $U_i$ is represented by
$g_i=(g_i^1,...,g_i^j,...,g_i^p)$ with 
$g_i^j=\sum_{s \in U_i}{x_s^j}$.

Finally, the prototype of each class been fixed, every element is assigned to a class according to its proximity to the prototype. The proximity is
measured by a classical distance between distributions (e.g. Euclidean distance):

$$d(x, y) = \sqrt{\sum_{j=1}^m (x_j - y_j)^2}, \, m \textrm{ is the number of modalities.}$$                                     

When there are dependencies between paths\footnote{For complete paths (option root=true and leaf=true), there are no embedded paths so (1) can be used.}, 
we replace the above formula by the following:

$$d(x, y) = \sqrt{\sum_{k=1}^p \sum_{j=1}^{m_k} (x_j^k - y_j^k)^2}$$
where $p$ is the number of variables, and $m_k$ is the number of  mo\-da\-li\-ties for the variable $k$

\section{Evaluation and metrics}
 Clustering evaluation is always a bit difficult, since, unlike classification, 
clustering is supposed to discover significant clusters whose number is not known in advance. 
However standard evaluation can be made on well known collection were some existing classification can be used as a reference.  
Since training sets are provided for the XML tracks we are able to evaluate our clustering approaches using them. 
We used different standard measures and compared their behavior when increasing the number of clusters and using different path lengths.  
We recall below the definition of the four metrics we use: F-measure, entropy, purity and corrected rand.

\begin{itemize}

\item The {\bf F-measure} proposed by Larsen and Aone~\cite{larsen99fast} combines the
precision and recall measures from information retrieval and treats each cluster as if it
were the result of a query and each class as if it were the desired
answer to that query. It is the  \emph{harmonic mean} between precision and recall. 

\item The {\bf Corrected Rand Index} has been proposed by Hubert and Arabie~\cite{hubert85} to compare two partitions. This measure can be used to compare the resulting clusters with an existing partition, or to compare two partitions resulting of different automatic
clustering.

\item {\bf Entropy}: it measures the class distribution of each cluster. 
The smaller the entropy value, the better the clustering solution. 
A perfect clustering solution would be the one that leads to clusters that contain documents from only a single class, 
in which case the entropy will be zero~\cite{zhao01criterion}.

\item {\bf Purity}: measures the percentage of documents in a cluster that belong to the largest class of documents in this cluster.
In general the larger the value of purity, the better the clustering solution~\cite{zhao01criterion}.
\end{itemize}

\section{Experiments with the INEX collections}

The INEX XML mining track provides a number of collections for evaluating clustering methods. 
Some of them consist only of document tree structures (structure-only collections), 
while the others correspond to XML documents with textual content (structure and content collections). 
The training sets consist of a subset of the documents in each collection, together with the class they belong to. 
As a consequence the expected number of clusters for each collection is known in advance. 
Below we give a short summary of the test collections we use for our experiments. 
Unfortunately we were not able to carry out all the experiments before the workshop, 
in particular because of the size of the structure and content collections.

\subsection{The IEEE collections}
From the standard IEEE collection used in the INEX ad-hoc retrieval
experiments, the XML mining track's organisers have derived two 
collections for XML mining, namely INEX-s (structure only) and INEX-cs (content and
structure). 
They preprocessed both collections in order to eliminate useless tags, 
as well as to remove information (the name of the Journal) that would
identify obviously the class the document belongs to.
For INEX-s, the clustering task is to identify the two classes that correspond, first to \emph{Transactions} Journals, second to other Journals. 
It is expected that the two types of Journals use different parts of the IEEE DTD and that articles could be
easily partitioned into the two classes. 

For INEX-cs, the clustering task, using both the structure and the content of the articles, is to identify 
the six classes proposed in Denoyer~\cite{denoyer04apprentissage} and built from the 18 existing Journals.

\begin{table}

\centering

\tabcolsep = 2\tabcolsep
\caption{Results for Inex-s (training collection) for path length 3 and 4, and cluster number set to 2 or 4}
\begin{tabular}{|c|c|c|c|r @{.} l|r @{.} l|r @{.} l|r @{.} l|}
\hline\hline

Path length & Root & Leaf & No. of Clusters & \multicolumn{2}{c|}{Fmeasure} & \multicolumn{2}{c|}{Corr.Rand} & \multicolumn{2}{c|}{Entropy} & \multicolumn{2}{c|}{Purity}\\ 

\hline

 3 & T & F & 2 & 0&662 &  0&098 & 0&755 & 0&663\\ \hline
 
 3 & F & T & 2 & 0&667 &  0&105 & 0&745 & 0&667\\ \hline
 
 \textbf{3} & \textbf{T} & \textbf{F} & \textbf{4} & \textbf{0}&\textbf{661} &  \textbf{0}&\textbf{423} & \textbf{0}&\textbf{185} & \textbf{0}&\textbf{963}\\ \hline
 
 3 & F & T & 4 & 0&549 &  0&005 & 0&728 & 0&682\\ \hline
 
 4 & T & F & 2 & 0&625 &  -0&044 & 0&871 & 0&650\\ \hline
 
 4 & F & T & 2 & 0&655 &  0&087 & 0&757 & 0&655\\ \hline
 
 4 & T & F & 4 & 0&542 &  0&208 & 0&457 & 0&857\\ \hline
 
 4 & F & T & 4 & 0&545 &  -0&001 & 0&737 & 0&675\\ 
\hline
\end{tabular}

\label{tab:results}

\end{table}

\subsection{The Movie database collections}
	
The MovieDB  corpus is a set of XML documents describing movies. It was built using the IMDB database. 
It contains 9643 XML documents. Each document is labelled by one thematic category which represents the genre of the movie in the original collection and one structure category. 
There are 11 thematic categories and 11 possible structure categories which correspond to transformations of the original data structures.
There are four resulting test collections for clustering based only on structure, and two test collections for clustering using both content and structure.

\section{Results}

Since training sets were provided, we use them to evaluate our approach before getting from the track organisers the official results on the test collections.
\subsection{IEEE structure collection}

\begin{figure}

\begin{minipage}{.45\linewidth}

\ifpdf
	\centering
	\includegraphics[width=5.5cm,height=4cm]{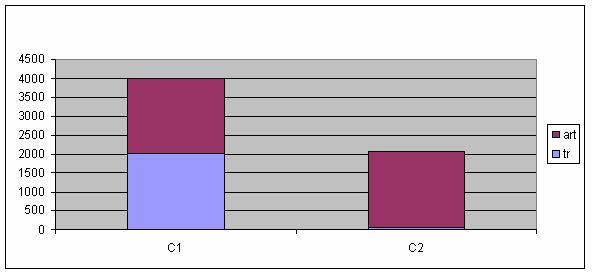}
\else
	\centerline{\psfig{figure=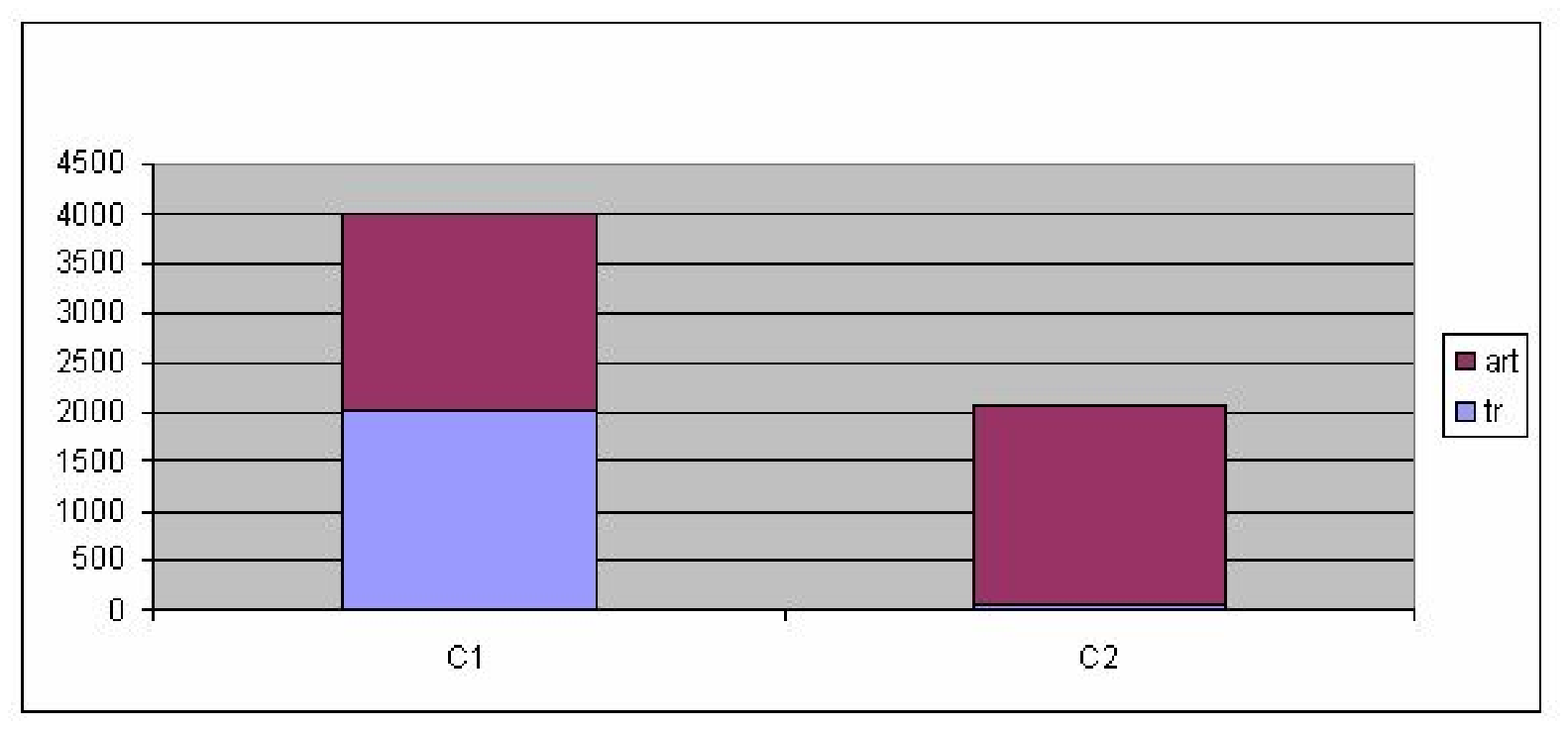,width=5.5cm,height=4cm}}
\fi
\caption{The repartition of classes \textit{Articles} and \textit{Transactions} on two clusters.}
\label{fig:inex-s_2clusters}

\end{minipage} \hfill
\begin{minipage}{.45\linewidth}

\ifpdf
	\centering
	\includegraphics[width=5.5cm,height=4cm]{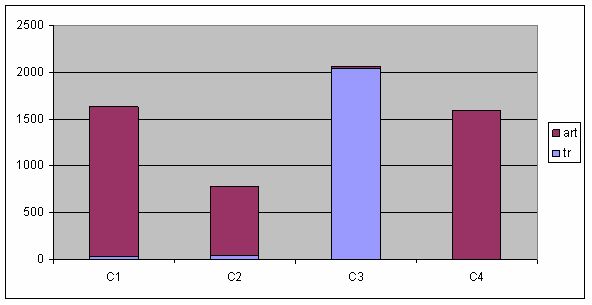}
\else
	\centerline{\psfig{figure=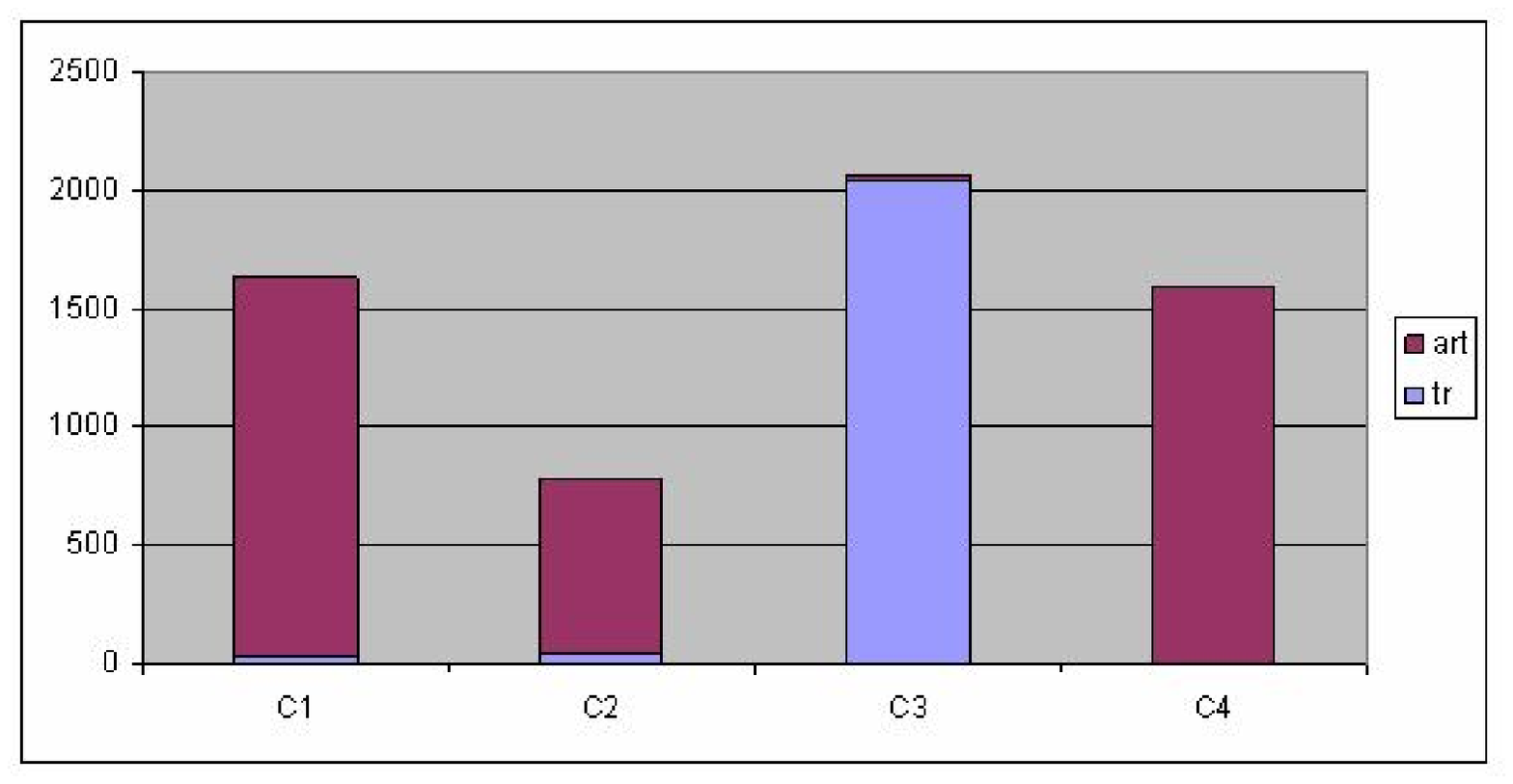,width=5.5cm,height=4cm}}
\fi

		\caption{The repartition of classes \textit{Articles} and \textit{Transactions} on four clusters.}

		\label{fig:inex-s_4clusters}

	\end{minipage}

 \end{figure}
 
We first test our approach with different path lengths either starting at the root or ending at the leaves.
We present only a few results in Table~\ref{tab:results}: The best values (especially entropy and purity) 
are obtained for documents represented by the set of paths of length 3 starting from the root. 
It must be noticed that it happens for four clusters, not the two that were expected.
There is nothing wrong with this result since there is no intrinsic reason why some articles would not have an overall structure more dissimilar to other articles than to {\it Transactions}. 

Fig.~\ref{fig:inex-s_2clusters} and Fig.~\ref{fig:inex-s_4clusters} shows the repartition of the two (resp. four) clusters on the two expected classes.
We can see that in the case of four clusters, one class (Transaction) maps quite closely to cluster 3, 
while the other three clusters contain mostly articles. 
We have not tried to analyse more deeply what could be the similarities between articles within these three clusters.

Then we sent two runs to the XML document mining track. 
The parameters we used and the official results are shown in Table~\ref{tab:officialresults}. 

\begin{table}

\centering

\tabcolsep = 2\tabcolsep
\caption{Official Results for inex-s (test collection) for two runs}
\begin{tabular}{|l|c|c|c|c|c|c|c|c|}
\hline

Run & Path-length & Root & Leaf & No. of & Micro  & Macro & Micro & Macro\\
    &        &      &   & Clusters & Entropy & Entropy &  Purity & Purity\\ 
\hline

 Run 1 & 3 & F & T & 2 &  0.744 & 0.627 & 0.663 & 0.627 \\ \hline
 
 Run 2 & 4 & T & F & 4  & 0.109 & 0.137 & 0.984 &  0.878\\ 
\hline
 
\end{tabular}

\label{tab:officialresults}

\end{table}
These results confirm the results with the training set that clustering in four clusters give better results than clustering in two clusters.

\subsection{MovieDB Structure Collections}

We did the same type of experiments for the four structured collections built from the Movie databases.
For each of the four collections we set the path length alternatively to 3 and 4, with either root paths or leaf-ending paths.
We cluster the documents into 9, 11 or 13 clusters respectively, the expected number of classes being 11.

\begin{table}

\centering

\tabcolsep = 2\tabcolsep
\caption{Results for m-db-s0 (training collection) for path length 3 and 4, and cluster number set to 9, 11 and 13}
\begin{tabular}{|c|c|c|c|r @{.} l|r @{.} l|r @{.} l|r @{.} l|}
\hline\hline

Path length & Root & Leaf & No. of Clusters & \multicolumn{2}{c|}{Fmeasure} & \multicolumn{2}{c|}{Corr.Rand} & \multicolumn{2}{c|}{Entropy} & \multicolumn{2}{c|}{Purity}\\ 

\hline

 3 & T & F & 9 & 0&541 &  0&370 & 0&286 & 0&632\\ \hline
 
 3 & F & T & 9 & 0&708 &  0&575 & 0&154 & 0&819\\ \hline
 
 3 & T & F & 11 & 0&509 &  0&357 & 0&285 & 0&640\\ \hline
 
 \textbf{3} & \textbf{F} & \textbf{T} & \textbf{11} & \textbf{0}&\textbf{642} &  \textbf{0}&\textbf{506} & \textbf{0}&\textbf{151} & \textbf{0}&\textbf{820}\\ \hline
 
 3 & T & F & 13 & 0&465 &  0&328 & 0&284 & 0&640\\ \hline
 
 3 & F & T & 13 & 0&595 &  0&473 & 0&151 & 0&819\\ \hline
 
 4 & T & F & 9 & 0&714 &  0&576 & 0&158 & 0&813\\ \hline
 
 4 & F & T & 9 & 0&714 &  0&576 & 0&158 & 0&813\\ \hline
 
 \textbf{4} & \textbf{T} & \textbf{F} & \textbf{11} & \textbf{0}&\textbf{663} &  \textbf{0}&\textbf{532} & \textbf{0}&\textbf{157} & \textbf{0}&\textbf{821}\\ \hline
 
 \textbf{4} & \textbf{F} & \textbf{T} & \textbf{11} & \textbf{0}&\textbf{663} &  \textbf{0}&\textbf{532} & \textbf{0}&\textbf{157} & \textbf{0}&\textbf{827}\\ \hline
 
 4 & T & F & 13 & 0&648 &  0&519 & 0&155 & 0&820\\ \hline
 
 4 & F & T & 13 & 0&648 &  0&519 & 0&155 & 0&820\\ 
\hline
\end{tabular}

\label{tab:results0}

\end{table}

Table~\ref{tab:results0} shows the measure values when clustering the training collection m-db-s0.
The results are always better when using leaf-ending paths over root paths, unless they are identical when the path length is set to 4.
The best value for the purity is obtained when clustering into 11 clusters, but the differences for other measures may not be all significant. 
We carried out similar experiments with the other collections, but there are not shown here for lack of space. 

Table~\ref{tab:officialresults-all} shows the official results for the four MovieDB collections for 11 clusters. 
As we can see, the quality of the results decreases with the increasing difficulty from m-db-0 to m-db-3.

\begin{table}

\centering

\tabcolsep = 2\tabcolsep
\caption{Official Results for Movie-DB (test collections) for two runs}
\begin{tabular}{|l|c|c|c|c|c|c|c|c|c|}
\hline

Coll. & Run & Path-lgth & Root & Leaf & Micro   & Macro   & Micro  & Macro & Mutual\\
     &     &           &      &      & Entropy & Entropy & Purity & Purity & Info\\ 
\hline
 m-db-0 & Run 1 & 3 & F & T &  0.732 & 0.841 & 0.203 & 0.136 & 1.823 \\ \hline
        & Run 2 & 4 & T & F &  0.732 & 0.841 & 0.203 & 0.136 & 1.823 \\ \hline
 m-db-1 & Run 1 & 3 & F & T &  0.688 & 0.804 & 0.326 & 0.226 & 1.528 \\ \hline
        & Run 2 & 4 & T & F &  0.707 & 0.835 & 0.256 & 0.144 & 1.690 \\ \hline
 m-db-2 & Run 1 & 3 & F & T &  0.688 & 0.758 & 0.296 & 0.209 & 1.592 \\ \hline
        & Run 2 & 4 & T & F &  0.458 & 0.501 & 0.487 & 0.446 & 1.139 \\ \hline
 m-db-3 & Run 1 & 3 & F & T &  0.623 & 0.714 & 0.316 & 0.238 & 1.545 \\ \hline
        & Run 2 & 4 & T & F &  0.553 & 0.636 & 0.527 & 0.438 & 1.044 \\ 
\hline
\end{tabular}

\label{tab:officialresults-all}

\end{table}

\begin{figure}

\ifpdf
	\centering
	\includegraphics[width=12cm,height=7.6cm]{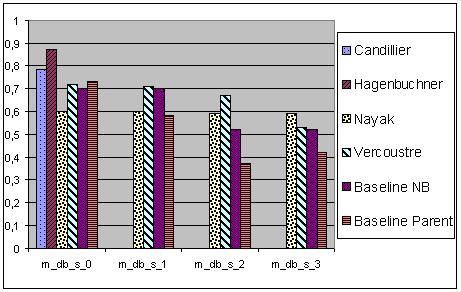}
\else
	\centerline{\psfig{figure=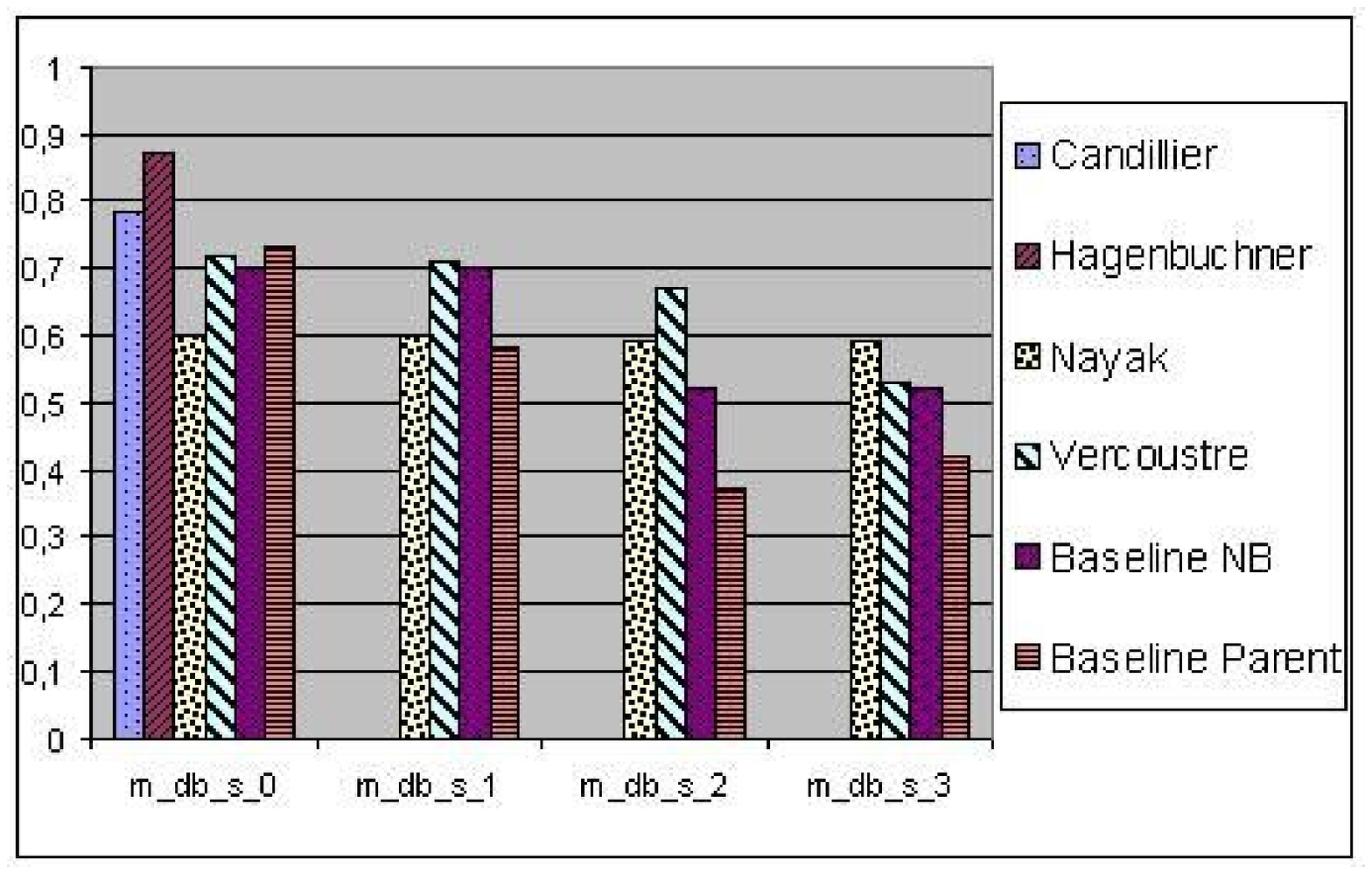,width=12cm,height=7.6cm}}
\fi
\caption{Comparison of clustering results for the Movie-db-s runs submitted to INEX 2005}
\label{fig:comparisons}
\end{figure}

We also include in Fig.~\ref{fig:comparisons}, the comparisons of our results with the runs submitted by other participants.
Our approach scores in the top-middle range of the four who have submitted results for the four collections.
Although Candillier~\cite{candillier05} and Hagenbuchner~\cite{Trentini05} submitted results only for Movie-db-s-0, 
their results are very promising. 

\subsection{Structure and Content Collection}

In Vercoustre and \textit{al.}~\cite{vfld06a} we had experimented the structure and content approach with two collections, including a small percentage of the INEX collection.
However we were not able to run our approach on the full Inex-cs collection, due to the large number of generated textual paths.
We did not experiment with m-db-cs collection but we expect that the same problem would occur.

Table~\ref{tab:vocabularysize} shows the number of different textual paths generated for different parameters, for 10\% of the collection. 

\begin{table}

\centering

\tabcolsep = 2\tabcolsep
\caption{Number of generated textual paths for 10\% of the inex-cs collection}
\begin{tabular}{|l|c|c|c|c|}
\hline

type & path-lgth & Root & Leaf & No. of distinct paths\\
\hline
 text       & 1 & F & T &  313078 \\ \hline        
 text+ tags & 1 & F & F &  340043 \\ \hline      
 leaf path  & 2 & F & T & 1271289 \\ \hline
 root path  & 3 & T & F &  367082 \\ \hline
 root path  & 4 & T & F &  387484 \\ 
\hline
\end{tabular}
\label{tab:vocabularysize}
\end{table}

The number of paths increases with the length of paths, and, for a fix path length, 
they are far more numerous for ending paths than for root paths since the tree is larger at the leaves than at the root.
Generating leaf paths is more costly and even overflows the generating program when the collection is too large.

\section{Conclusion}

In this paper we proposed to represent XML documents by a set of their paths generated according to a range of parameters. We evaluated our approach on some of the collections proposed by the INEX XML Mining track and we were relatively successful on the structure-only collections.
  
However, we have not managed to cluster the full structure-and-content collections, due to the large size of the generated vocabulary. We are thinking of reducing the vocabulary by using the TF/IDF frequency of terms in each specific path, rather than the frequency of textual paths in a document and the collection respectively.

In both cases, structure and structure-and-content, it could also be beneficial to reduce the space dimension before clustering, for example by using Principal Component Analysis like in Liu and \textit{al.}~\cite{liu04pca}.

%
%

\bibliographystyle{abbrv}
\bibliography{INEXpaper}

\end{document}